\documentclass[%
 reprint,
amsmath,amssymb,
 aps,
]{revtex4-2}

\usepackage{graphicx}
\usepackage{dcolumn}
\usepackage{xcolor}
\usepackage{ragged2e}
\usepackage{bm}
\usepackage{float}
\usepackage{xspace}
\newcommand{\Poincare}{Poincar\'e\xspace}

\begin{document}

\preprint{APS/123-QED}

\title{Evidence of Chaotic Mixing in the Alveolar Region of the Lung}

\author{Prabhash Kumar}
\author{Prahallada Jutur}%
 


\author{Anubhab Roy}
 
\author{Mahesh V. Panchagnula}
\email{mvp@iitm.ac.in}
\affiliation{%
Department of Applied Mechanics, Indian Institute of Technology Madras,
Chennai 600036, India 
}%


\date{\today}

\begin{abstract}

We report an experimental and numerical investigation to study the role of asymmetry in the expansion-contraction of the acinar wall on the particle transport in the acinus. We model the acinar flow feature using a T-section by appropriately matching the dimensionless numbers to that in the acinus of healthy human subjects. We show that asymmetry in the expansion-contraction process (quantified by $\phi$) is required for chaotic advection. We show the stretch and fold process leading to chaos for a range of $\phi$ and scaled oscillation frequency $Sr$. We show a regime map in this generalize $\phi$ and $Sr$ space and show that most mammalian lungs fall at the boundary of chaotic regime.

\end{abstract}

\maketitle


\textit{Introduction} - The airways that lead to the lungs from the nasal passage go through bifurcations at multiple levels, finally ending in the smallest lung feature known as alveoli \cite{west2012respiratory}. Symmetrical branching models \cite{weibel1963morphometry,horsfield1968morphology,horsfield1971models,phillips1994diameter} describe the lung airways as comprising of a total of 23 generations, where after each generation, a bronchus bifurcates into two daughter bronchioles. The bronchioles become narrower, shorter, and more numerous as they move deep into the lungs. The fluid-particle dynamics attain a unique transport feature after each bifurcation due to the wide spectrum of Reynolds numbers ranging from $O(>10^3)$ in the trachea to $O(>10^{-3})$ in the acinar region \cite{kleinstreuer2010airflow,sznitman2013respiratory,verbanck2011gas,sznitman2021revisiting}. Understanding the dynamics of \textit{small} inhaled particulate matter in the acinus is crucial to understand the propagation of deadly respiratory diseases like SARC-CoV-2 and SARS \cite{donaldson2005combustion,anderson2012clearing,londahl2014measurement,scheuch2020breathing,leung2021transmissibility} as well as to improve the efficiency of target-oriented pharmaceutical aerosols deposition \cite{patton2007inhaling,azarmi2008targeted}. The alveolar ducts and alveoli in the acinar region occupy $90\%$ of the total lung volume and actively participate in the gas exchange process \cite{weibel1984pathway,sznitman2013respiratory}. A study of fluid-particle dynamics in the acinus is non-trivial due to the modelling difficulties in the sub-millimetre-sized alveolus and the coupling between the complex respiratory flows and intrinsic motion of particles \cite{kleinstreuer2010airflow}.

For decades, researchers believed that the flow features in the acinus are kinematically reversible and trivially simple, as the flow Reynolds number is much less than unity~\cite{davidson1972flow,cinkotai1974fluid,ultman1985gas}. This belief persisted until the seminal experiment of \citet{heyder1988convective}. Through a non-diffusive aerosol bolus experiment on 17 healthy male subjects, the authors reported that `convective' mixing occurs in the lung parenchyma and is, therefore, not only a feature of upper respiratory airways. Although the reason for this increased mixing in the acinus was not clear and incorrectly attributed as convective, the authors believed that the reason could be one or more of the following: (a) airway and alveoli geometry, (b) asymmetries in the inspiratory and expiratory velocity profiles, (c) disparities in air velocity due to unequal airway dimension in the same generation and (d) asymmetrical expansion and contraction in alveoli. Later, Tsuda and the co-authors~\cite{tsuda1995chaotic,butler1997effect,tsuda1995chaotic,tippe2000recirculating,haber2000shear} reported that the complexity of acinar flow occurs due to the time-dependent motion and alveolated structure of the acinar wall. The authors reported the existence of a saddle point near the mouth of the alveolar opening for low values of the ratio of alveolar flow to the ductal flow, pointing to a possible chaotic flow situation.

It is evident from these previous studies that ultrafine sub-micron particles could reach the acinus, avoiding the geometrical barriers in the upper lung generation. However, the mode by which they can mix and deposit in the acinus is not quite settled science. In particular, impaction is not active since particle inertia can be ignored at this scale. The occurrence of increased mixing in the acinar flows under these circumstances is perplexing and, at the same time, important as they contribute to pulmonary disease inception. Several articles appeared over the decades investing in the acinar airflow characteristics and the effects of intrinsic particle motion on particle deposition in the lung acinus~\cite{tsuda1994effects,tsuda1994effects,tsuda1995chaotic,darquenne1996two,ma2011aerosol,ma2012aerosol,sznitman2009respiratory,sznitman2010visualization,henry2012simultaneous,sera2015numerical}. Most of these studies consider symmetric airflow geometry and the self-similar expansion of the lung while others \cite{miki1993geometric, berg2010flow} report the presence of asymmetric motion of alveoli during expansion and contraction and report~\cite{tsuda1999acinar, kumar2009effects, chhabra2010flow, ciloglu2020numerical} the significant role of irreversible acinar motion on the transport of particles in higher airway generations. The question of the origin of chaotic advection in alveolar flows is still not answered.

\begin{figure}[!htb]
\centering
\includegraphics[width = 0.45\textwidth]{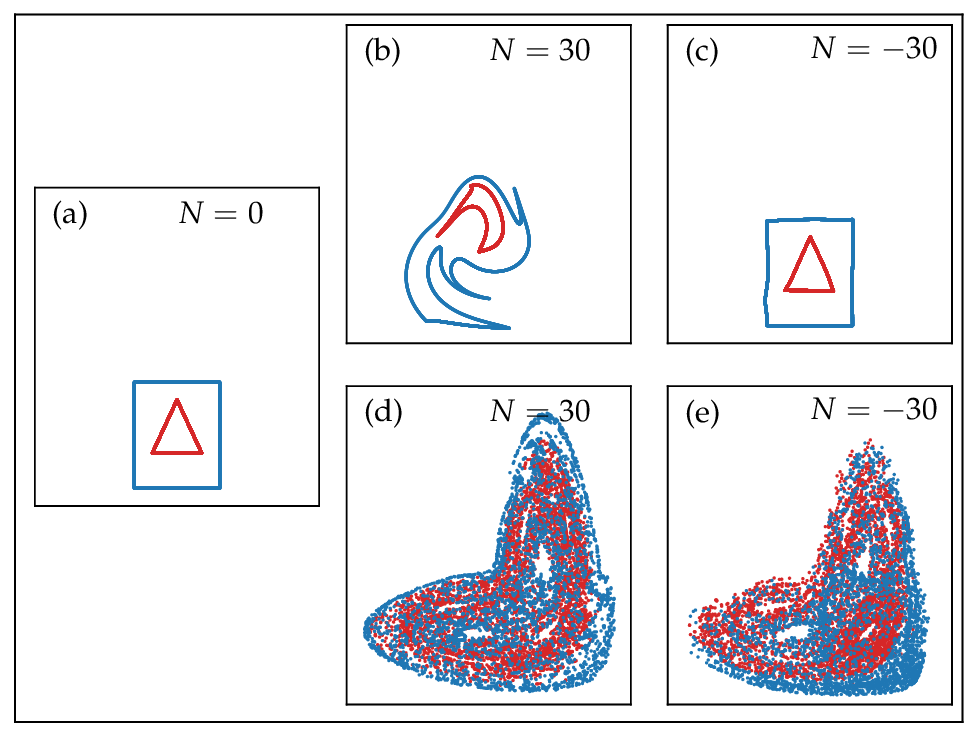}

\caption{\label{fig:Reversibility} The $30$ cycles forward and $30$ cycles backward map of tracers to differentiate the non-chaotic and chaotic advection. The inset (a) plots the initial location of tracers arranged in two different configurations: circle and triangle. The insets (b) and (c) show the forward and backward maps of non-chaotic advection, respectively, and the insets (d) and (e) show the forward and backward maps of tracers advection of chaotic case. The initial location for backward simulation is the final location of forward simulation, and the initial location of particles at $N=0$ is identical for both chaotic and non-chaotic cases.}
\end{figure}

In this letter, we report an experimental and numerical investigation to understand the inception of chaotic mixing and the role of asymmetric expansion and contraction of the acinar wall on particle transport. We show conclusive evidence of chaotic mixing in an alveolus as being responsible for particle transport at that scale. In figure~\ref{fig:Reversibility}, we show the cases at two different parametric states. We run one forward simulation with two different initial particle locations and find the location of particles at $30$ cycles ahead. Then, we run the backward simulation for $30$ cycles by reversing the flow field direction and treating the final location of the forward simulation as the initial location of this case. We find that beyond a critical combination of parametric state $Sr$ and $\phi$ ($Sr$ is the dimensionless oscillation frequency and $\phi$ is an asymmetric parameter), the final location of the backwards simulation and the initial location of the forward simulation are identical, which is not the case otherwise. As shown in the figure, we find a chaotic, disordered state of tracers below the critical values of these parameters. Interestingly, the final particle locations in the case of chaotic motion are independent of the initial locations of the particle, which is preserved even when the flow field is reversed.
We will briefly describe our experimental setup and simulation details and discuss the parametric condition beyond which particle motion becomes chaotic. 

\textit {Experimental setup and simulations details} - We employ an experimental setup, which generates the viscous flow features of the lung acinus (more details are available in the supplemental material). This setup employs two independently driven pistons in a T-junction simulating an expanding/contracting alveolus. These pistons execute the time-periodic motion. The difference in flow magnitude in both the pistons determines an asynchrony parameter $\phi$, where $V_{L}/V_{R}=(\sec{\phi}-\tan{\phi})$. Here, $V_{L}$ and $V_{R}$ are the velocities of left-side and right-side pistons, respectively. Similarly, we solve the equations for unsteady Stokes flow with physics-controlled finer meshing with around $1080$ domain and $151$ boundary elements in  COMSOL\textsuperscript{\textregistered} to obtain a computed fluid flow field. Using the computed flow field, we solve for the trajectory of tracers.

\begin{figure}[!htb]
\includegraphics[width = 0.45\textwidth]{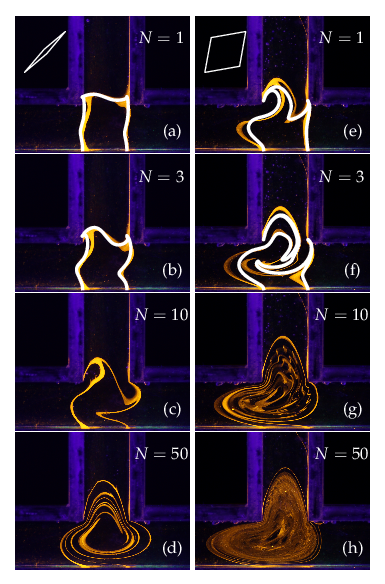}
\caption{\label{fig:Phase-locked_Image_At_Different_phi} The \Poincare 
 maps of experimental images at $N=1,\;3,\;10,\;\text{and}\;50$. Figures (a)-(d) correspond to the case of $\phi=10^{\circ}$ at $Sr=0.5$ and (e)-(h) correspond to $\phi=70^{\circ}$ at $Sr=0.5$. The insets of figures (a) and (e) indicate the displacement (left-side Vs right-side) curve of pistons over one oscillation of respective cases. The white patterns in figures (a), (b), (e), and (f) are from computer simulations for the respective cases, and all the patterns in yellow are from experiments.}
\end{figure}

\textit{Results} - We inject dyed silicone oil at the T-junction in an inverted U shape for experimental visualization. For quantification purposes, we predominantly use flow visualization with time-lapse imaging separated by one period of oscillation. To understand the mixing nature, we capture several phase-locked time-lapse images (typically for about $500$ cycles). Initial experiments reveal that the Reynolds number has no significant role as long as it is much smaller than unity $(Re\ll 1)$. In addition, the convective flow patterns remain the same even for different $Re$ conditions as long as Strouhal number $Sr=fL/U$ is constant. Here, $f$ oscillation frequency, $L$ is the characteristics dimension of the T-junction, and $U$ is the characteristics velocity. Therefore, we consider the effect of $\phi$ and $Sr$ as the relevant parameters to the current study. The dye pattern is advected, stretched, folded and regularly returns to the initial position at $\phi=0^{\circ}$. In this case, the phased-locked images are almost identical for all $Sr$, indicating no mixing. However, the differential piston velocity at even a small non-zero $\phi$ exhibits an additional rotation in the flow field, breaking the periodicity. We observe significant aperiodic deformation for even $\phi\geq 5^{\circ}$. We show the phase-locked images of the tracer dye for $\phi=10^{\circ}$ and $\phi=70^{\circ}$ for $Sr=0.5$ in Fig. \ref{fig:Phase-locked_Image_At_Different_phi}. For $\phi=10^{\circ}$, we observe the buckling in the dye, which folds on each other in a nice regular fashion with increasing the number of cycles $N$. At the end of $50$ cycles, a structure is formed with various layers of dye lamina stacked on top of each other, having a finite gap between them. However, the structure for the higher $\phi=70^{\circ}$ case is significantly complex. It is comprised of irregular and buckled set of layers which merge into each other even as early as after $10$ cycles. For $N=50$, the structure is completely well-mixed on the scale of this image. In this case, we observe the appearance of extra features in layers of dye on increasing $N$, which makes it irregular. Hence, increasing the $\phi$ beyond a critical value breaks both the periodicity and the regularity of the dye deformation, significantly increasing mixing.

While performing the different sets of experiments, we observe that once periodicity is broken at a small non-zero $\phi$, even decreasing $Sr$ adds irregularities in the deformation. Based on our experiments, we report that a small non-zero $\phi$ is required to break the symmetry, and the combination of high $\phi$ and low $Sr$ adds further complexity to the mixing process. Moreover, the forward and backward simulation in Fig.~\ref{fig:Reversibility} shows the existence of a chaotic attractor for specific values of $\phi$ and $Sr$. Now, we formally quantify these distinctions by calculating the line length of the dye. Beyond an initial few cycles, calculating the line length of the dye pattern from the experimental images is troublesome because of overlap and continuity; hence, we use data from computer simulations to quantify the line length and the resulting parameters. We use tracer particle simulation and calculate a summation of inter-particle distances over all the particles to find the total line length. 
\begin{equation}
    L(t) = \sum_{i=0}^{i=n}||\mathbf{x}_{i+1}(t)-\mathbf{x}_i(t)||
\end{equation}
where $n$ is the number of particles and $||.||$ is the $L_2$ norm of the vector. The increase in the pattern's line length is crucial to understanding mixing. We compare our simulation results with experiments in the insets of figure~\ref{fig:Phase-locked_Image_At_Different_phi} (a) and (e), where deformation from simulation closely matches with the phased-locked images. The comparison plot eliminates the role of experimental artefacts like piston misalignment and time-delay in obtaining phased-locked images.

\begin{figure}[!htb]
\includegraphics[scale=0.6]{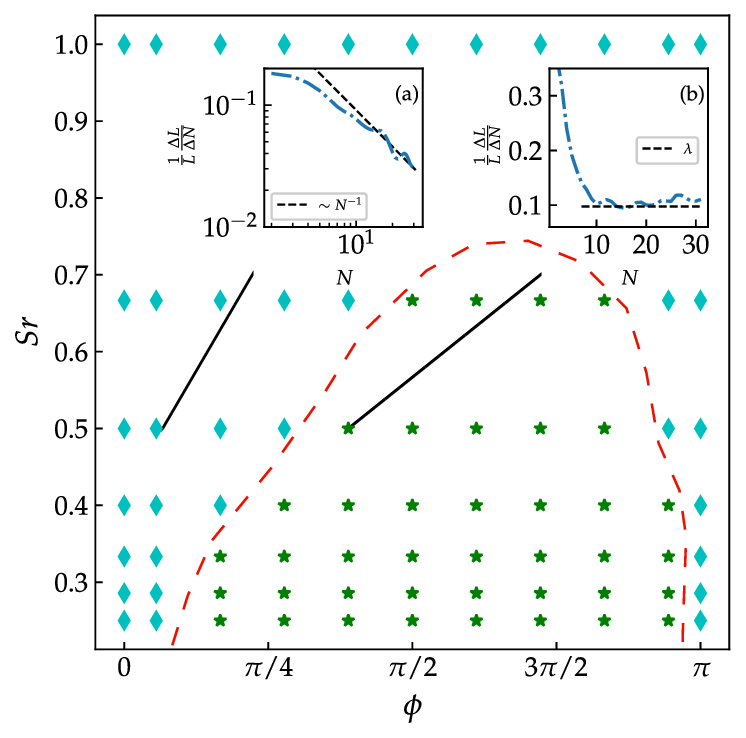}
\caption{The regime plot of $Sr$ Vs $\phi$ indicates the regions of non-chaotic and chaotic advection of tracers. Here, the $\blacklozenge$ symbol represents the non-chaotic advection, and $*$ indicates the chaotic advection. The insets (a) and (b) plots $\frac{1}{L}\frac{\Delta L}{\Delta N}$ as function of $N$ for a non-chaotic case $(\phi=10^{\circ},\;Sr=0.5)$ and chaotic case $(\phi=70^{\circ},\;Sr=0.5)$, respectively. $\lambda$ is the Lyapunov exponent of the corresponding chaotic case. The red dashed line is the separating boundary of chaotic and nonchaotic regions.}
\label{fig:ContourPlot}
\end{figure}

\begin{figure}[!htb]
\includegraphics[scale=0.5]{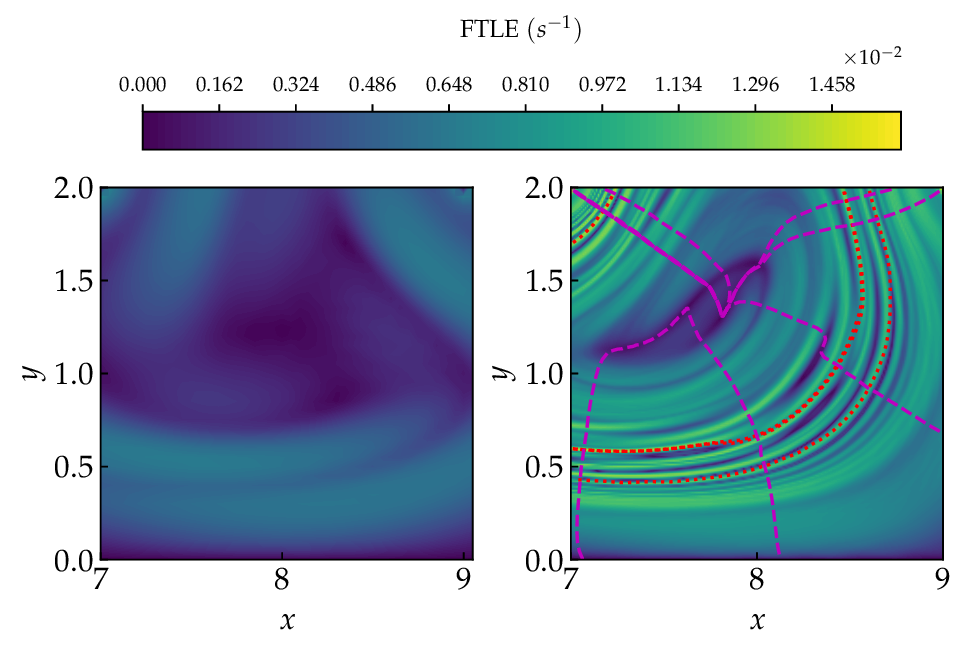}
\caption{Finite Time Lyapunov exponent (FTLE) field corresponding to (a) nonchaotic advection and (b) chaotic advection. In plot (b), the dotted line (red) indicates the location of repelling LCS, and the dashed line (magenta) indicates the location of attracting LCS.}
\label{fig:FTLE}
\end{figure}

In order to quantify the occurrence of chaos, we calculate the line length for different cases—the line length scales in two different ways as a function of the number of cycles. The increase in line length is exponential for $70^{\circ}\leq\phi\leq 150^{\circ}$ at $Sr=0.5$ indicating the onset of chaos and the span of $\phi$ for exponential growth increasing on reducing $Sr$, implying that at $Sr=0.25$ the exponential growth occurs for $30^{\circ}\leq\phi\leq 170^{\circ}$. In contrast, it is slower than exponential for any $\phi$ at high $Sr$. We indicate a regime map with chaotic and non-chaotic behaviour of particle motion in the figure~\ref{fig:ContourPlot} by calculating the Lyapunov exponents from the simulation data. The particle motion at the two extreme ends of the $\phi$ axis ($\phi=0^{\circ}$ and $\phi=180^{\circ}$) is always non-chaotic and periodic irrespective of $Sr$. As we move down the $Sr$ axis, we find a transition behaviour followed by Chaotic behaviour. The span of chaotic behaviour on the $\phi$ axis increases on reducing $Sr$ and is constrained by the extreme points. We calculate the quantity $\frac{1}{L}\frac{\Delta L}{\Delta N}$ from the line length to represent exponential and sub-exponential growth. $\frac{1}{L}\frac{\Delta L}{\Delta N}$ should scale as $N^{-1}$ for the sub-exponential case, which appears as a straight line on the log-log plot as shown in the inset (a) of figure~\ref{fig:ContourPlot}. In this case, the intercept of the ordinate axis gives information on the exponent of power-law growth. For the exponential growth, $\frac{1}{L}\frac{\Delta L}{\Delta N}$ attain a constant value, which should be the Lyapunov exponent, as shown in the inset (b) of figure~\ref{fig:ContourPlot}. \\  
To further affirm the chaotic behaviour led by the exponential growth in the line length, we detect the most influential material lines, in this case, along which the stretching/shrinking is locally the largest. The Finite Time Lyapunov exponent (FTLE) $(\sigma)$ provides the average exponent of this stretch/shrink ~\cite{shadden2005definition,haller2015lagrangian},
\begin{equation}
    \sigma_{t_0}^{T}(\mathbf{x}) = \frac{1}{|T|}\ln{\sqrt{\lambda_{max}(\Gamma)}},
\end{equation}
where $\Gamma = \left[\nabla\phi_{t_0}^{T}\right]'\nabla\phi_{t_0}^{T}$ is the right Cauchy-Green deformation tensor. $\phi_{t_0}^{T}$ is the Lagrangian description of fluid motion known as a flow map, which gives the final state $(\mathbf{x}, T)$ after a time of interest $T$ given the initial state $(\mathbf{x}_0, t_0)$. The symbol $(\;'\;)$ represents the transpose of the matrix. $\lambda_{max}(\Gamma)$ is the maximum eigenvalue of the $\Gamma$ representing the maximum stretching occurring over the time interval $T$. We calculate the field of FTLE at $t_0=0$ on a grid of particles at the T-junction over the time interval of 10 oscillation cycles. Fig.~\ref{fig:FTLE} shows the FTLE fields corresponding to non-chaotic (a) and chaotic cases (b). In the chaotic case, the stretch rate at the most influential location is around an order larger than that in the nonchaotic case. The material lines (in this case) along the sharp ridges of FTLE fields are the candidates for the Lagrangian coherent structures (LCS), also known as transport barriers in the flow field. We use the variational theory~\cite{farazmand2012computing} and extract the LCS corresponding to the chaotic advection by computing the tensor lines. The tensor lines tangent to the eigenvector field associated with the maximum eigenvalue of $\Gamma$ are the repelling LCS. Similarly, the attracting LCS is the locus of points tangent to the eigenvector field associated with the minimum eigenvalue of $\Gamma$. We indicate the most prominent repelling (dot line) and attracting LCS (dashed line) in figure~\ref{fig:FTLE}~(b) 

\textit{Conclusion} - We report an experimental investigation coupled with numerical simulation to identify the role of asymmetric acinar wall motion on particle transport in the acinus. The flow features depend on a motion asymmetric parameter of acinar walls $\phi$ and the Strouhal number $Sr$ (scaled oscillation frequency). The advection of tracers undergoes time-periodic motion for $\phi=0^{\circ}\; \text{and}\;180^{\circ}$ irrespective of $Sr$. A small offset value of $\phi$ from $0^{\circ}$ and $180^{\circ}$ breaks the time-periodicity of the motion. The aperiodic advection turns chaotic on increasing $\phi$ at a fixed $Sr$ and decreasing $Sr$ at fixed $\phi$. We have quantified the non-chaotic and chaotic advection by showing the algebraic and exponential growth in the line length of the pattern formed by the tracers and identified the region of chaotic advection in the parametric space $Sr$ and $\phi$. By computing the FTLE field and extracting the Lagrangian coherent structures, we show that the stretch rate at a most influential location in chaotic is around an order larger than that in a nonchaotic case and indicated the most prominent coherent structures in chaotic case. In general, one can not control the asymmetric acinar wall movement during breathing. However, for given $\phi$, we conclude that slow $\&$ deep breathing results in better mixing particles in the deep lung.

\bibliography{apssamp}
\bibliographystyle{apsrev4-1}
\end{document}


\preprint{APS/123-QED}

\title{Supplemental material}
\maketitle
\section{Details of experimental apparatus}
\begin{figure}[!htb]
\includegraphics[width = 0.45\textwidth]{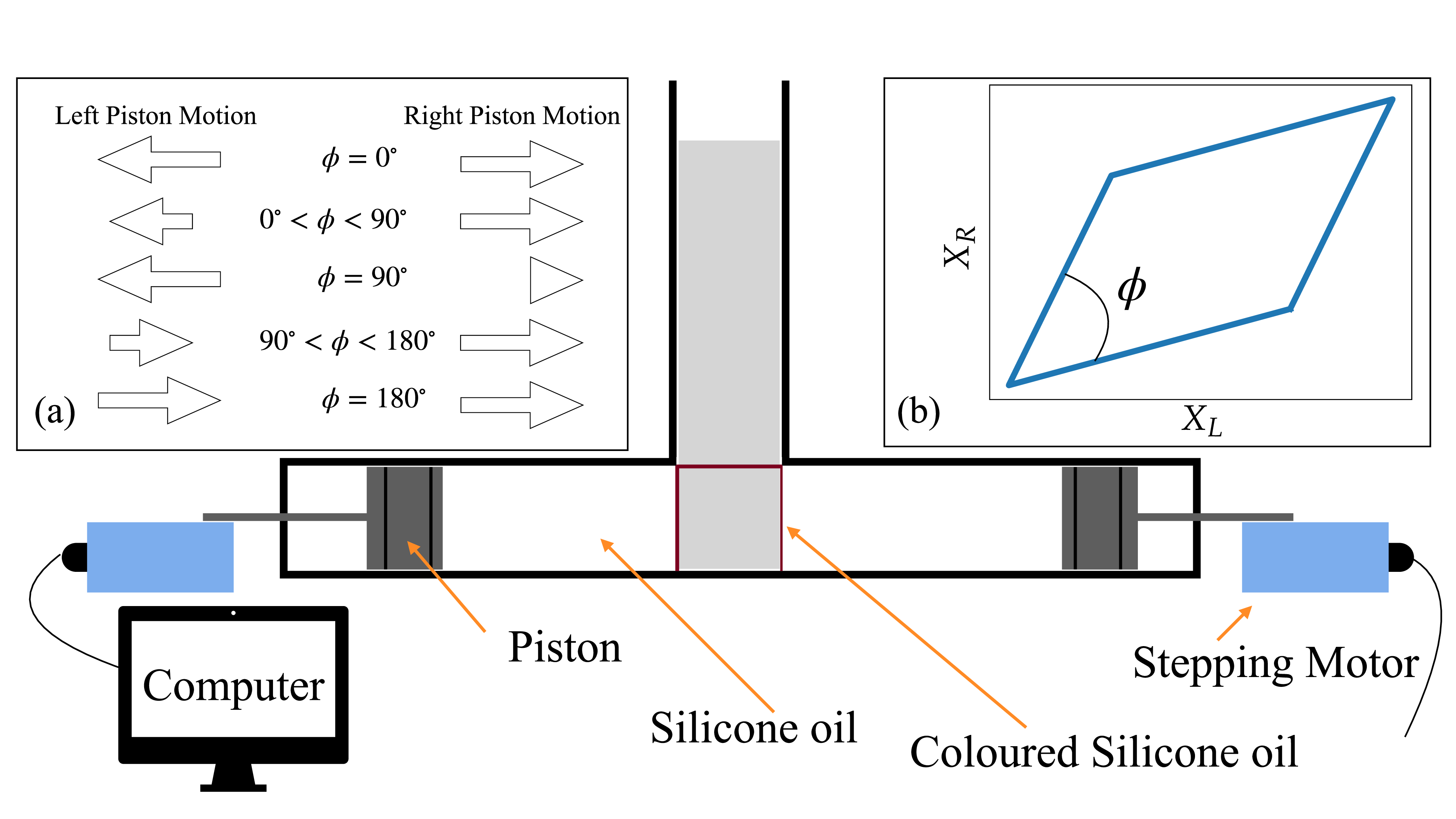}
\caption{\label{fig: Experiment_schematic} Schematic of the experimental setup (T-section with stepper motors). The top left inset box (a) depicts the piston movement at different values of the asymmetry parameter $\phi$, where the arrow length represents the magnitude of piston velocity. The inset on the top right (b) shows the displacement curve of the pistons.}
\end{figure}

We model the acinar flow feature using the T-section of a square cross-section of side $20$ mm. The horizontal arms connect to the syringe-piston systems, and the vertical arm opens to the atmosphere, as shown in Fig.~\ref{fig: Experiment_schematic}. 
Two programmable independent stepper motors drive the syringe-piston system to pump the fluid in the T-section. We appropriately chose the Harvard Pump 11 Elite to cater to the low flow rate requirements and achieve the low Reynolds number flow, $Re=UL/\nu<1$. Here, $U$ and $L$ are characteristic velocity and length scales, and $\nu$ is kinematic fluid viscosity. Pistons rhythmically infuse and withdraw the fluid to mimic the inhalation and exhalation processes. During an oscillation, the pumps withdraw/infuse the fluid into the T-section with a constant velocity, depending on the asymmetry parameter $\phi$ pistons infuse or withdraw. The following relation between the piston velocities:
\begin{equation}
    V_1/V_2=(\sec{\phi}-\tan{\phi})
\end{equation}
governs the magnitude and direction of the velocity of the pistons to introduce the acinar geometric asymmetry into the system. Here, $V_1$ and $V_2$ are the velocities of the pistons. The left panel (a) in figure~\ref{fig: Experiment_schematic} shows the direction and magnitude (represented by the length of the arrow) of pistons' velocity for a given $\phi$. For $\phi=0^{\circ}$, both the pistons withdraw or infuse at the same time with equal and opposite velocities (symmetric case); in the first oscillation's quarter, both pistons withdraw and both infuse for the next two quarters followed by a final withdrawal by both the pistons in the last quarter. For $0^{\circ}<\phi<90^{\circ}$, the magnitudes of pistons' velocity are unequal. Still, they withdraw or infuse similarly as $\phi=0^{\circ}$ and keep switching velocity in each quarter to ensure kinematic reversibility at the end of each cycle. For $\phi=90^{\circ}$, one piston moves while the other remains at rest. In the first quarter, the right position withdraws, and the left piston stands still. Then, the left piston infuses, and the right piston stands still in the next quadrant. In the third quadrant, the right piston infuses, the left piston rests, and in the final quadrant, the left piston withdraws, and the right piston remains at rest. For $90^{\circ}<\phi<180^{\circ}$, both the pistons move with unequal velocities in the same direction, like in the first two quarters, the right piston withdraws, and the left piston infuses and then changes the velocity direction in last two quarters. For $\phi=180^{\circ}$ is again the symmetric case but is different than the $\phi=0^{\circ}$, as fluid mass does not stretch much in this case because the pistons move in the same direction with equal velocities; as a result the fluid height in vertical arm remain unchanged during the motion. The right panel (b) in figure~\ref{fig: Experiment_schematic} shows the piston displacement curve over an oscillation in the plane of the right-side piston displacement, $X_R$ and the left-side piston displacement $X_L$. The angle $\phi$ of this curve is essentially the geometric asymmetry, and the area within this curve measures the degree of flow asymmetricity, which should be zero for $\phi=0^{\circ}$ and $\phi=180^{\circ}$. We use the silicon oil of dynamic viscosity $10000$ cSt as the working fluid to match the necessary low Reynolds number conditions $(Re = 0.008)$ of the alveolar ducts of the $23^{\text{rd}}$ generation. With the help of two sets of UV lights of a wavelength of $395$ nm and placing a coloured silicone oil (coloured with oil-soluble fluorescent powder) at the T-junction, ensuring it is sufficiently below the free surface, we visualise the flow patterns.
\section{Details on numerical simulations}
In simulations, we use the identical T-junction geometry with identical dimensions as in experiments and solve the equations for unsteady Stokes flow with physics-controlled finer meshing with around $1080$ domain and $151$ boundary elements in COMSOL\textsuperscript{\textregistered} to obtain a computed fluid flow field. Taking the flow data from COMSOL\textsuperscript{\textregistered}, we do the spatiotemporal interpolation and solve for trajectories of tracer particles using the following equation
\begin{equation}\label{EquationOfMotion}
    \frac{d\mathbf{x}}{dt}=\mathbf{u}(\mathbf{x}, t),    
\end{equation}
where $\mathbf{x}$ is particle position and $\mathbf{u}(\mathbf{x}, t)$ is the interpolated flow field in the T-section domain.